\documentclass[amsmath,superscriptaddress,reprint]{revtex4-2}
\usepackage[english]{babel}
\PassOptionsToPackage{english}{babel}
\usepackage{lineno}
\usepackage[utf8]{inputenc}
\usepackage{color}
\usepackage{graphicx,hyperref}
\usepackage{booktabs}
\usepackage{soul}
\usepackage{float}

\hypersetup{
    bookmarks=false,         
    unicode=false,          
    pdftoolbar=false,        
    pdfmenubar=true,        
    pdffitwindow=false,     
    pdfstartview={FitH},    
    pdftitle={Dual relaxation oscillations in a Josephson junction array},    
    pdfauthor={Higginbotham group},     
    pdfsubject={},   
    pdfcreator={},   
    pdfproducer={}, 
    pdfkeywords={}, 
    pdfnewwindow=true,      
    colorlinks=true,       
    linkcolor=black,          
    citecolor=blue,        
    filecolor=magenta,      
    urlcolor=blue           
}

\addto\captionsenglish{}

\begin{document}
\selectlanguage{english}

\title{Dual relaxation oscillations in a Josephson junction array}

\author{S. Mukhopadhyay}
\affiliation{IST Austria, Am Campus 1, 3400 Klosterneuburg, Austria}
\author{D.A. Lancheros-Naranjo}
\affiliation{IST Austria, Am Campus 1, 3400 Klosterneuburg, Austria}
\author{J. Senior}
\affiliation{IST Austria, Am Campus 1, 3400 Klosterneuburg, Austria}
\author{A.P. Higginbotham}
\email{ahigginbotham@uchicago.edu} 
\affiliation{The James Franck Institute and Department of Physics, University of Chicago, Chicago, Illinois 60637, USA} 
\affiliation{IST Austria, Am Campus 1, 3400 Klosterneuburg, Austria}

\begin{abstract}
We report relaxation oscillations in a one-dimensional array of Josephson junctions.
The oscillations are circuit-dual to those ordinarily observed in single junctions. 
The dual circuit quantitatively accounts for temporal dynamics of the array, including the dependence on biasing conditions.
Injection locking the oscillations results in well-developed current plateaux.
A thermal model explains the relaxation step of the oscillations.
\end{abstract}

\maketitle

\section{Introduction}
\label{sec:intro}
\begin{figure*}
	\centering
	  \includegraphics[width=0.8\textwidth]{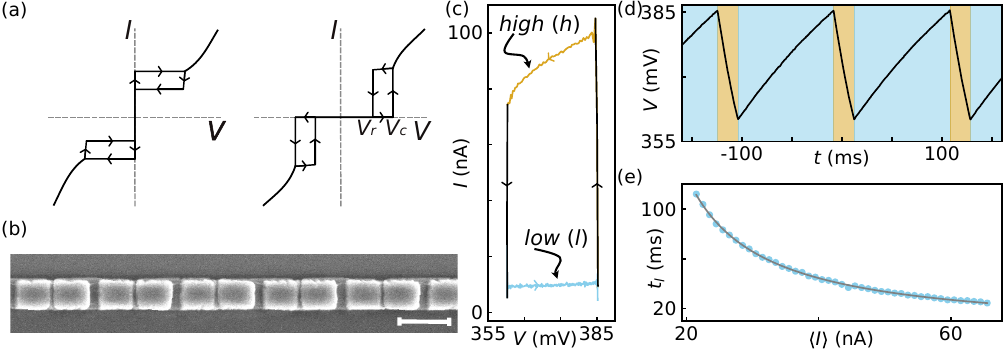}
		\caption{
		Phenomenology of dual relaxation oscillations.
		(a) Schematic current-voltage characteristic of a current biased Josephson junction (left) and a voltage biased Josephson array (right). 
		Arrows indicate orbits of relaxation oscillations for the ordinary (left) and dual (right) cases. 
		Note that the ordinary and dual case have opposite chirality orbits.
		$V_r$ is the relaxation voltage.
		$V_c$ is the critical voltage.
		(b) Scanning electron micrograph of the studied device. 
		The scale bar represents $1~\mathrm{\mu m}$.
		(c) Parametric plot of the measured current $I(t)$ versus the measured voltage $V(t)$, consistent with the hysteretic behavior illustrated in (a, right).
		Regions of low current $l$ and high current $h$ are indicated.
		Four arrowheads are added to illustrate chirality of the current-voltage orbit.
		(d) Voltage $V$ measured as a function of time $t$.
		The blue regions represent states $l$ and the orange regions represent states $h$ (as illustrated in (c)). 
		Plots in (c) and (d) are obtained when the array is biased with a current of $23.5~\mathrm{nA}$.
		(e) Extracted dwell time $t_{l}$ (in state $l$) as function of the biasing current $\langle I \rangle$.
		The grey curve shows fit to Eq.(\ref{eq:tl}).
		The bias current is denoted as $\langle I \rangle$ because it fixes the time-averaged current flowing in the circuit.
		}
	  \label{fig:iv_intro}
\end{figure*}

A current-biased Josephson junction is hysteretic in measured voltage.
When voltage biased, such a junction can undergo relaxation oscillations controlled by an inductive time constant, generating voltage plateaux when phase locked~\cite{vernon_relaxation_1968,whan_effect_1995}. 
In the electrically dual scenario, of interest here, a voltage-biased device exhibits hysteretic behavior in measured current [Fig.~\ref{fig:iv_intro}(a)].
When current biased, it undergoes relaxation oscillations controlled by a capacitive time constant, generating current plateaux when phase locked.

Current-voltage dualities in superconducting circuits have long been of interest as, under some circumstances, they can reflect phase-charge dualities in the circuit Hamiltonian.
A visible case is the Schmid transition \cite{schmid_diffusion_1983,bulgadaev_phase_1984} in a high-impedance environment, where a single Josephson junction is predicted to transition into a charge-localized insulating state, dual to the phase-localized superconducting state.
Early transport experiments observed insulating behavior in qualitative agreement with theory \cite{kuzmin_coulomb_1991,yagi_phase_1997,penttila_superconductor-insulator_1999}.
More recently, a lively experimental \cite{murani_absence_2020,kuzmin_observation_2023} and theoretical \cite{morel_double-periodic_2021,masuki_absence_2022,carles_absence_2023} body of work has emerged scrutinizing this transition.
Signatures of Bloch oscillations \cite{averin_bloch_1985}, dual to Shapiro steps, have also been observed \cite{lehtinen_coulomb_2012}, and in recent breakthroughs have now been reported with part-per-thousand precision \cite{shaikhaidarov_quantized_2022,crescini_evidence_2023}.
Of particular relevance to this work, current hysteresis \cite{agren_kinetic_2001,vogt_one-dimensional_2015,cedergren_insulating_2017} and dynamics \cite{lotkhov_cooper_2007} have been observed in voltage-biased Josephson arrays, constituting a current-voltage dual to the single junction case.
The nature of the hysteresis and of the dynamics is unclear: are they low-frequency signatures of Bloch physics, or are they of a different origin? 

Here, we observe relaxation oscillations in a current-biased one dimensional array of Josephson junctions.
These oscillations are well described by a circuit model, dual to the ordinary Josephson relaxation oscillations~\cite{vernon_relaxation_1968}. 
Injection locking these oscillations results in current plateaux.
The relaxation step is found to obey a characteristic self-consistent relation, suggesting that it is governed by overheating effects.

Our platform for studying dual relaxation oscillations is a chain of $1217$ Al/AlOx Josephson junctions fabricated by a standard Dolan bridge process using electron-beam lithography [Fig.~\ref{fig:iv_intro}(b)].
Measurements are performed using standard transport techniques, with filtering and methods described in Ref.~\cite{mukhopadhyay_superconductivity_2023}.
The arrays we study are known to exhibit weak supercurrent features visible at low bias \cite{mukhopadhyay_superconductivity_2023}.
At large biases these features are not apparent, and the system exhibits a resistive low-bias branch with a hysteretic critical and relaxation voltage [App.~\ref{app:i_v}].
Similar voltage hysteresis has been observed in fully insulating arrays \cite{agren_kinetic_2001,vogt_one-dimensional_2015,cedergren_insulating_2017}.
The hysteretic I-V characteristic of the chain gives rise to relaxation oscillations in certain ranges of bias currents.

Relaxation oscillations emerge when an external current bias $\langle I \rangle$ is imposed in the unstable regime ($\mathrm{11.5~nA<}~\langle I \rangle~\mathrm{<69.5~nA}$).
The oscillations are characterized by a periodic, spiking voltage signal, reminiscent of a charge-discharge cycle of a capacitor [Fig.~\ref{fig:iv_intro}(d)].
We have verified that changing room-temperature biasing circuitry does not qualitatively alter the oscillations.
Parametrically plotting the instantaneous measured current and voltage reveals a relaxation orbit with distinct low-current ($l$) and high-current ($h$) regions [Fig.~\ref{fig:iv_intro}(c)].
Correlating states $l$ and $h$ to the time-domain voltage oscillations, it can be seen that in the $l$ state the voltage steadily increases, until the system quickly transitions to the $h$ state, where the voltage steadily decreases.
Based on the chirality of the current-voltage orbit, the observed relaxations are unambiguously of the dual type.

The time spent in each state depends on the external biasing conditions.
Plotting the low-current state dwell time $t_l$ at different bias currents $\langle I \rangle$ shows that $t_l$ smoothly decreases as $\langle I \rangle$ is increased [Fig.~\ref{fig:iv_intro}(e)].
We have checked that the complementary dwell time in the high-current state, $t_h$, has the opposite dependence on bias [App.~\ref{app:th_fit}], and that the time average of the instantaneous current over a cycle corresponds to the externally applied bias.

\begin{figure}
	\centering
	\includegraphics[width = 0.45\textwidth]{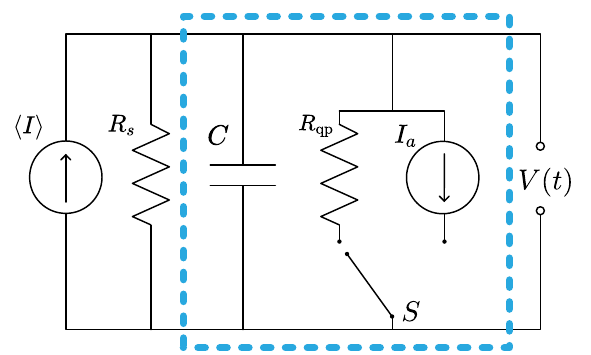}
	\caption{{Dual circuit model.}
	$\langle I \rangle$ is the applied current bias. 
	$R_s$ denotes the output impedance of the current biasing unit, and $C$ denotes the net capacitance associated with the circuit.
	The dashed blue box represents the cryostat.
	Switch $S$ changes the circuit between states $l$ and $h$.
	In the low current state $l$ (blue section of the I-V characteristic of Fig.~\ref{fig:iv_intro}(c)), the array is in its quasiparticle branch with resistance $R_{\text{qp}}$. 
	In the high current state $h$ (orange section of the I-V characteristic of Fig.~\ref{fig:iv_intro}(c)), the array acts as a current source $I_{a}$.
	Voltage $V\mathrm{(t)}$ indicates the measured voltage.
	}
	\label{fig:dual_circuit}
\end{figure}

\section{Proposed circuit model}
\label{sec:circuit_model}
For a quantitative understanding of the dual relaxation oscillations, we introduce an effective circuit model consisting of a parallel RC circuit fed by a current source $\langle I \rangle$ [Fig.~\ref{fig:dual_circuit}]. 
This model is dual to the circuit for regular Josephson relaxation oscillations~\cite{vernon_relaxation_1968}, see App.~\ref{app:duality}.
The circuit has two states, corresponding to the $l,h$ oscillation states, controlled by a switch.

In the $l$ state the Josephson array is represented by a resistance $R_\mathrm{qp}$, and the total circuit resistance $R_{||}$ is the parallel of the device resistance and the current-source output impedance $R_s$, $R_{||} = (R_{s}^{-1}+R_{\text{qp}}^{-1})^{-1}$.
The device voltage $V$ approaches the steady-state $\langle I \rangle R_{||}$ with a time constant $\tau_{l} = R_{||} C$, according to 
\begin{equation}
	\label{eq:vl}
	V(t) = V_{r}+\left(\langle I\rangle R_{||} - V_{r}\right)\left(1-\exp(-t/\tau_{l})\right).
\end{equation}
However, $V$ does not reach its steady-state value.
Rather, when $V=V_c$ the switch flips to the $h$ state, where the array acts as a current source $I_a$.
The output voltage then approaches the steady-state value $(\langle I \rangle - I_a) R_s$ with time constant $\tau_{h} = R_{s}C$, according to
\begin{equation}
	\label{eq:vh}
	V(t) = V_{c} + \left(\left(\langle I\rangle - I_{a}\right)R_{s} - V_{c}\right)\left(1-\exp(-t/\tau_{h})\right).
\end{equation}
When $V=V_r$ the switch flips back to the $l$ state, and the cycle repeats.

The dynamics can be solved for the dwell time in each state.
The dwell time in state $l$ is
\begin{equation}
	\label{eq:tl}
	t_{l} = \tau_{l}\log\left(1+\frac{V_r-V_c}{V_c - \langle I\rangle R_{||}}\right), 
\end{equation}
and the dwell time in state $h$ is
\begin{equation}
	\label{eq:th}
	t_{h} =  \tau_{h} \log\left(1 + \frac{V_{c} - V_{r}}{V_{r} - (\langle I\rangle - I_{a})R_{s}}\right).
\end{equation}

To compare with experiment, parameters $V_{c}$ and $V_{r}$ are fixed by identifying the switching voltages in Fig.~\ref{fig:iv_intro}(d).
$R_\mathrm{qp}=44.4~\mathrm{M \Omega}$  is fixed from a linear fit to wide bias current-voltage data [App.~\ref{app:i_v}].
The dwell time data in Fig.~\ref{fig:iv_intro}(e) are then fit to Eq.~(\ref{eq:tl}), yielding $R_s=460~\mathrm{M \Omega}$ and $C$.
The fit capacitance $C=53.2~\mathrm{nF}$ is close to the value from filters in the measurement setup ($47~\mathrm{nF}$)~\cite{mukhopadhyay_superconductivity_2023}.
The remaining free parameter, $I_a$, is found by fitting $t_{h}$ versus $\langle I \rangle$ to Eq.~(\ref{eq:th}) with all other parameters held fixed [App.~\ref{app:th_fit}].
The fit value $I_a=87.2~\mathrm{nA}$ is compatible with typically observed current values in the $h$ state ($\approx 90~\mathrm{nA}$).

\section{Comparison of experiment and dual circuit model}
\label{sec:comparison}
\begin{figure}
	\includegraphics[scale=0.95]{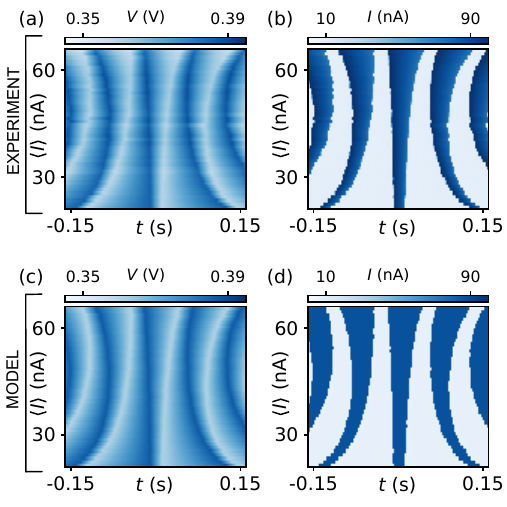}
  	\caption{
	 {Dual relaxation oscillations.}
	 Measured relaxation oscillations of the voltage $V$ (a) and current $I$ (b) as a function of time $t$ and bias current $\langle I \rangle$.
	 Predicted $V(t,\langle I \rangle)$ (c) and $I(t,\langle I \rangle)$ (d).
	 Voltage is calculated from Eqs.~(\ref{eq:vl}-\ref{eq:vh}).
	 Current in the $l$ state is calculated from $V(t)/R_\mathrm{qp}$, taking $V(t)$ from Eq.~(\ref{eq:vl}).
	 Current in the $h$ state is simply $I_a$.
	 }
	\label{fig:tdomain}
\end{figure}

Having fixed all parameters within the dual Josephson model, it is now possible to directly compare it with the full behavior of dual relaxation oscillations.
Measuring the time resolved voltage oscillations at different current biases reveals the oscillation time period to be shorter at intermediate biases ($\mathrm{30~nA>}$ $\langle I \rangle$ $\mathrm{>60~nA}$) than at extremal biases [Fig.~\ref{fig:tdomain}(a)].
The measured current shows a similar behavior [Fig.~\ref{fig:tdomain}(b)], with distinct regions of low ($l$) and high ($h$) current [App.~\ref{app:model_iv_t}, Fig.~\ref{fig:iv_t}(b)].
Voltage and current calculated from the dual Josephson circuit, within the bias range $\mathrm{21.5~nA<}~\langle I \rangle~\mathrm{<65.5~nA}$, have excellent qualitative agreement with the experimental observations [Fig.~\ref{fig:tdomain}(c),(d)].
This provides strong evidence that our oscillations are circuit-dual to the ordinary Josephson relaxation oscillations.

\section{Phase-locked oscillations}
\label{sec:lock-in}
\begin{figure*}
	\centering
	\includegraphics[width = 0.8\textwidth]{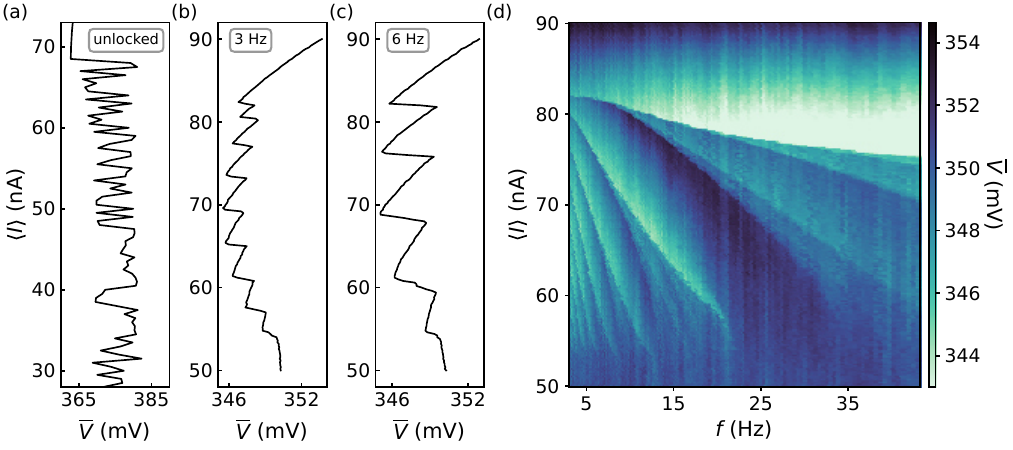}
  	\caption{
	 {Current steps in voltage locked traces.}
	 (a) Bias current $\langle I \rangle$ as a function of measured voltage $\overline{V}$.
	 $\overline{V}$ represents the voltage recorded after averaging over a few cycles of $V$ versus $t$ data.
	 (b) Current bias as a function of measured voltage with a locking frequency of 3 Hz.
	 (c) Same measurement as in (b) with a locking frequency of 6 Hz.
	 To improve the signal to noise ratio, data in (b) and (c) are plotted after averaging over 50 measurement runs.
	 (d) Voltage $\overline{V}$ measured as a function of bias current $\langle I \rangle$ and locking frequency $\mathrm{f}$.
	 }
	 \label{fig:locking}
\end{figure*}

To gain further insight into the temporal dynamics of the oscillations, we inject a sinusoidal locking signal with amplitude and frequencies comparable to the voltage oscillations and measure the average voltage over an integer number of time periods.
In contrast to the unlocked state where the dual Josephson relaxation oscillations appear to generate noise [Fig.~\ref{fig:locking}(a)], in the presence of the locking excitation a series of well-developed current plateaux emerge [Fig.~\ref{fig:locking}(b)].
Doubling the locking frequency approximately halves the number of observed current plateaux [Fig.~\ref{fig:locking}(c)].
Measuring the phase-locked average voltage versus current traces for a range of locking frequencies result in a fan of current plateaux in the current-frequency plane [Fig.~\ref{fig:locking}(d)].

The I-V characteristics in the locked state are also quantitatively different from the unlocked state.
The amplitude of current plateaux in the locked state is $\sim 5~\mathrm{mV}$, which is several times smaller than the voltage noise amplitude in the unlocked state $\sim 20~\mathrm{mV}$ [Fig.~\ref{fig:locking}(a-c)].
In addition, the voltage measured in locked state is shifted by $\sim 25~\mathrm{mV}$ when compared to the unlocked state.
It is also interesting to note that the current steps are only observed for high bias currents $\langle I \rangle \gtrsim 55~\mathrm{nA}$.

For ordinary Josephson relaxation oscillations, phase locking is known to generate voltage plateaux~\cite{vernon_relaxation_1968}.
We have observed the electrically dual oscillations, and the corresponding dual current plateaux.
Up to this point our analysis has been at a circuit level, phenomenologically encoding dynamics at $V_c$ and $V_r$ with a switching circuit element.
This analysis is sufficient for demonstrating that we have observed a circuit dual to Josephson relaxation oscillations, but leaves open the underlying cause.
Below, we argue that thermal effects play an important role in the switching physics.

\section{Thermal model}
\label{sec:thermal}

\begin{figure*}
	\includegraphics[width=0.85\textwidth]{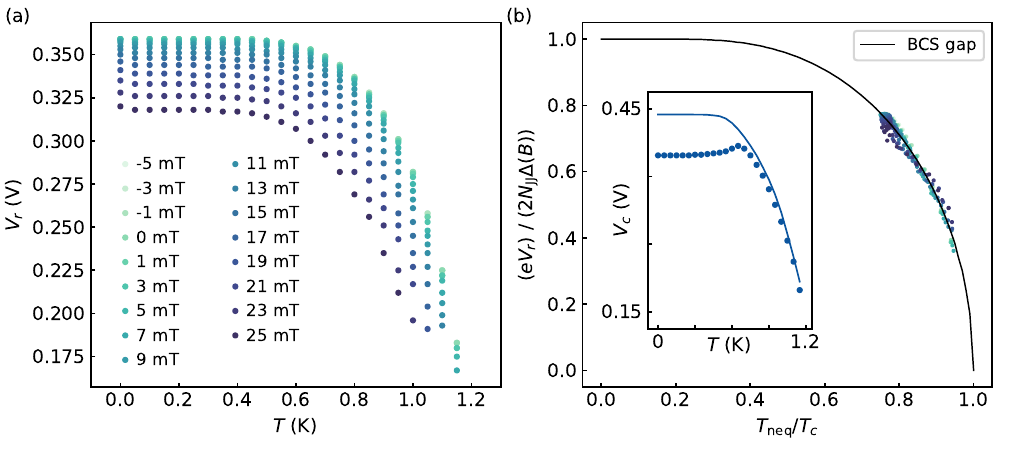}
  	\caption{
	 {Role of heating in generating hysteresis.}
	 (a) Relaxation voltage $V_r$ as a function of temperature $T$ at various magnetic fields.
	 Magnetic field is applied in plane of the chip, perpendicular to the axis of the Josephson chain. 
	 (b) Plotting normalized $V_r$ against the normalized, non-equilibrium effective temperature (${T_\mathrm{neq} / T_c}$) collapses the data in (a), see Eq.~(\ref{eq:vr_normed}).
	 Solid black line is the normalized BCS gap function.
	 Inset shows temperature dependence of critical voltage $V_c$.
	 The solid blue line is the expected equilibrium evolution $e V_{c} = 2 N_\mathrm{JJ} \Delta(T)$.
	 }
	 \label{fig:thermal_model}
\end{figure*}

To gain insight into the microscopic mechanism governing the relaxation physics, we have measured the relaxation voltage $V_r(B,T)$ in a stable, voltage-biased configuration as a function of in-plane magnetic field $B$ and cryostat temperature $T$.
$V_r$ is insensitive to small temperature changes, but decreases dramatically in the vicinity of Aluminum's critical superconducting temperature $T_{c,0}\sim 1.2~\mathrm{K}$ [Fig.~\ref{fig:thermal_model}(a)], confirming the superconducting origin of the relaxation oscillations.
Applied magnetic field decreases $V_r$ and moves its temperature dependence down to lower scales.

A simple picture of overheating, along the lines suggested in Ref.~\cite{agren_kinetic_2001}, explains the evolution of $V_r(B,T)$.
The fundamental assumption is that the Josephson array switches at a voltage proportional to the superconducting gap.
In the $h$ state, Joule heating reduces the superconducting gap, resulting in $V_r < V_c$.
We make a few simplifying assumptions to arrive at a tractable model.
Relaxation is assumed to occur when each junction is biased by twice the superconducting gap $\Delta$, $e V_{r} = 2 N_\mathrm{JJ} \Delta( B, T_\mathrm{neq} )$.
The gap is assumed to be spatially uniform, but to depend on a single non-equilibrium effective temperature $T_\mathrm{neq}$, found from the condition that Joule heating balances cooling due to electron-phonon coupling,
$
I_r V_r = \Sigma \mathcal{V} (T_\mathrm{neq}^5 - T^5),
$
where $I_r$ is the current at the relaxation point, $\Sigma$ is the electron-phonon coupling, $\mathcal{V}$ is the device volume, and $T$ is the cryostat temperature.
At non-zero magnetic fields, the gap is approximated as retaining a BCS form $\Delta_\mathrm{BCS}$ \footnote{we have checked numerically that this is adequate for small fields}, and as being scaled from its zero-field value according to
$
\Delta(B, T) = \Delta(B) \Delta_\mathrm{BCS}( T/T_c )/\Delta_0,
$ where $\Delta(B) = \Delta_0 \sqrt{1-(B/B_c)^2}$, and $T_c = T_{c,0} \sqrt{1-(B/B_c)^2}/\sqrt{1+(B/B_c)^2}$ based on a two-fluid model with weak quadratic pair breaking \cite{tinkham_introduction_1996}.
Here $\Delta_0$ is the zero-temperature, zero-field gap and $B_c$ is the critical field.
Combining the expression for $\Delta(B)$ with the criteria for relaxation gives 
\begin{equation}
	\label{eq:vr_normed}
	\frac{e V_r}{2 N_\mathrm{JJ} \Delta(B)} = \frac{\Delta_\mathrm{BCS}( T_\mathrm{neq}/T_c )}{\Delta_0}.
\end{equation}
Given measured $I_r$ and $V_r$, Eq.~(\ref{eq:vr_normed}) constitutes a self-consistency test for thermal relaxations. If obeyed, the scaled relaxation voltage (left-hand side), is equal to the normalized BCS gap function at a self-consistent, non-equilibrium temperature $T_\mathrm{neq}$ (right-hand side).

Equation (\ref{eq:vr_normed}) is fit to the experimentally measured $I_r$ and $V_r$ with $\Sigma$, $B_c$ and $T_{c,0}$ as free parameters.
Scaling the data with the best-fit parameters results in a reasonable collapse of the data onto the normalized BCS gap function, as expected [Fig.~\ref{fig:thermal_model}(b)].
The best-fit parameter values agree with physical expectations: $T_{c,0}=1.26~\mathrm{K}$ and $\Sigma = 0.5~\mathrm{nW/(\mu m^3 K^5)}$ are typical of aluminum \cite{agren_kinetic_2001,wellstood_hot-electron_1989,kauppinen_electron-phonon_1996}, and $B_c = 68~\mathrm{mT}$ is compatible with the expected critical field of the thicker islands in our chain \cite{mukhopadhyay_superconductivity_2023}.
Both the collapse of the data and the agreement of fit parameters with expectations suggests that the relaxation step is dominated by thermal effects.

Interestingly, however, the critical voltage is not well described by a thermal model.
$V_c$ increases slightly as temperature is increased [Fig.~\ref{fig:thermal_model}(b), inset], 
which disagrees with the expected dependence for equilibrium switching, and cannot be explained by overheating alone.
Near the Aluminum critical temperature, behavior consistent with the thermal model is recovered for $V_{c}$.

\section{Summary and Outlook}
\label{sec:summary}

We have uncovered thermal effects in the superconducting state of a one dimensional Josephson array.
Our thermal model sheds light on the origin of hysteresis in voltage biased arrays; the non-equilibrium effective temperature $\mathrm{T_{neq}}$ resulting from Joule heating scales the superconducting gap such that the chain relaxes to a low-current state at a voltage $V_r$, lower than the critical voltage $V_c$.
Based on our analysis, Joule overheating results in $T_\mathrm{neq} \sim 0.9~\mathrm{K}$ at base cryostat temperature.

Our proposed dual circuit model explains the temporal relaxation dynamics of a current-biased Josephson array.
Phase locking to the voltage oscillations produces current plateaux, which are dual to the voltage plateaux observed in ordinary Josephson relaxation oscillations \cite{vernon_relaxation_1968}.

Overall, this work emphasizes thermal load management in the superconducting state of a Josephson chain.
In contrast to our earlier work studying thermal effects in equilibrium \cite{mukhopadhyay_superconductivity_2023}, this studies overheating at large applied bias.

\section*{Acknowledgements}
We gratefully acknowledge support from the MIBA machine shop and Nanofabrictation Facility at IST Austria.
Work was supported by Austrian FWF grant P33692-N (S.M., J.S. and A.P.H.), the European Union's Horizon 2020 Research and Innovation program under the Marie Skłodowska-Curie Grant Agreement No. 754411 (J.S.), and a NOMIS foundation research grant (A.P.H.).

\appendix

\section{Wide bias I-V characteristics}
\label{app:i_v}

\begin{figure}[H]
	\centering
	\includegraphics[scale=0.7]{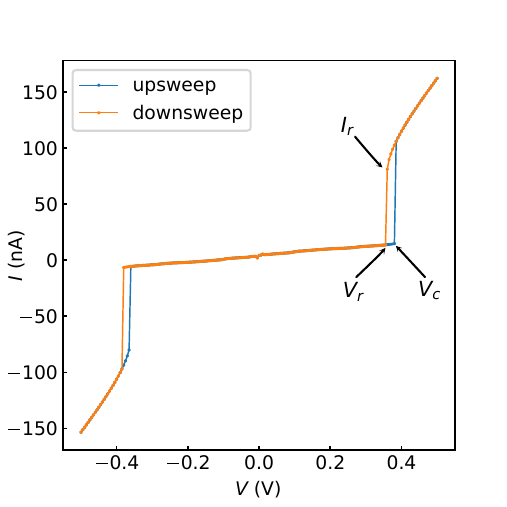}
  	\caption{
	 {Current-voltage characteristics of the Josephson chain.}
	 $V_c$ is the critical voltage.
	 $V_r$ is the relaxation voltage.
	 $I_r$ is the current at the relaxation point.
	 }
	\label{fig:i_v}
\end{figure}
In the main text, we discussed the voltage oscillations resulting from the bi-stability in a current-biased device.
Fig.~\ref{fig:i_v} shows the voltage-biased curves and the region of bi-stability (between $V_r$ and $V_c$), resulting from voltage upsweep and down-sweep on the same device.
Note that the bistable region for positive voltage biases is the same as the one described in Fig.~\ref{fig:iv_intro}(c).
This serves as a consistency check between the voltage biased and current biased curves. 

\section{Temporal response predicted from model}
\label{app:model_iv_t}

\begin{figure}[H]
	\includegraphics[scale=0.95]{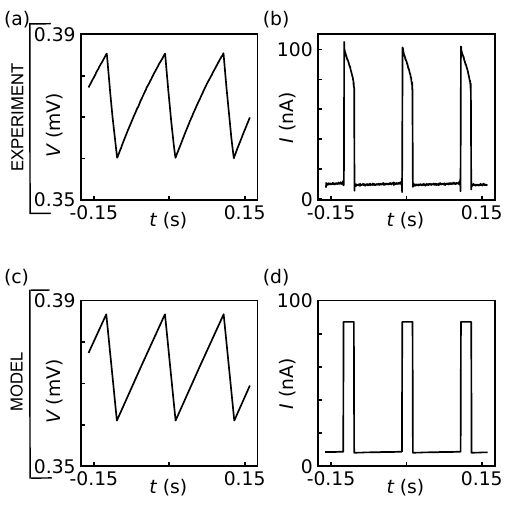}
  	\caption{
	 {Cuts at $\langle I \rangle = \mathrm{23.5~nA}$ from Fig.~\ref{fig:tdomain}.}
	 (a) Relaxation oscillations of the voltage $V$ measured as function of time $t$.
	 (b) Measured current $I$ versus time $t$ corresponding to the region of relaxation oscillations. 
	 The period of low current $t_{l}$ and high current $t_{h}$ are clearly distinguishable.
	 (c) Calculated voltage $V$ versus time $t$ using Eqs.(\ref{eq:vl} and \ref{eq:vh}), based on the circuit model of Fig.~\ref{fig:dual_circuit}. 
	 (d) Calculated current $I$ versus time $t$ within the theoretical model.
	 }
	\label{fig:iv_t}
\end{figure}
In Sec.~\ref{sec:comparison} of main text, we compare plots from experiment and the model over two-dimensional parameter space. 
In Fig.~\ref{fig:iv_t}, we explicitly compare the voltage and current, versus time, by taking cuts from the data in Fig.~\ref{fig:tdomain}.
The calculated current in Fig.~\ref{fig:iv_t}(d) is constant in the $h$ state, which is expected from the model due to absence of any output impedance in the current generator $I_a$.

\section{Fit of high state dwell times}
\label{app:th_fit}

\begin{figure}[H]
	\centering
	\includegraphics[scale=0.7]{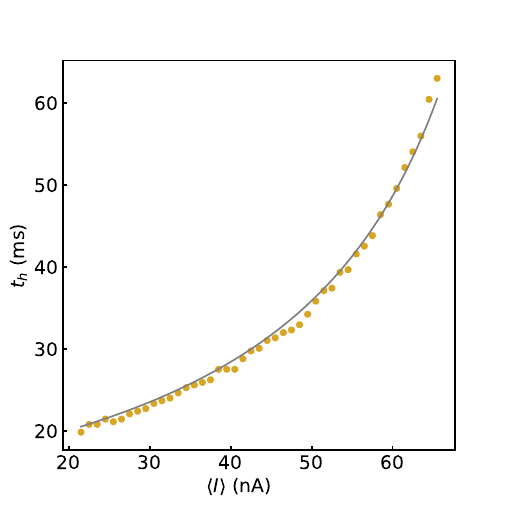}
  	\caption{
	 Extracted dwell time $t_h$ as a function of biasing current $\langle I \rangle$.
	 The grey curve shows a single-parameter fit to Eq.~(\ref{eq:th}) for $I_a$.
	 }
	\label{fig:th}
\end{figure}
We mentioned earlier that the circuit parameters in the proposed model are derived from fitting Eqs.~(\ref{eq:tl},\ref{eq:th}) to the dwell time data.
Fig.~\ref{fig:th} shows the plot of high state dwell times versus biasing current, complimenting the plot in Fig.~\ref{fig:iv_intro}(e).
At higher bias currents $\langle I \rangle$, the system ends up spending most of its time in state $h$, as evident from the asymmetry in hysteretic region of Fig.~\ref{fig:i_v}.
Hence, the dwell time in state $h$ increases with higher bias current, while the dwell time in state $l$ decreases [Fig.~\ref{fig:iv_intro}(e)].

\section{Duality with Josephson relaxation oscillations}
\label{app:duality}

It is interesting to note that the expressions for $t_l$ and $t_h$ (Eqs.~\ref{eq:tl},\ref{eq:th}) are analogous to the expressions worked out for ordinary Josephson relaxation oscillations~\cite{vernon_relaxation_1968}, albeit with $V_{r} = 0$. 
The duality substitutions are $I_{c}R\rightarrow V_{c}$ and $V_{g}\rightarrow I_{a}R_{s}$. 
Depending on the state ($h$ or $l$), $L/R \rightarrow R_{s}C$ or $R_{||} C$, and $V_{0}\rightarrow \langle I\rangle R_{s}$ or $\langle I\rangle R_{||}$. 


\begin{thebibliography}{25}%
	\makeatletter
	\providecommand \@ifxundefined [1]{%
	 \@ifx{#1\undefined}
	}%
	\providecommand \@ifnum [1]{%
	 \ifnum #1\expandafter \@firstoftwo
	 \else \expandafter \@secondoftwo
	 \fi
	}%
	\providecommand \@ifx [1]{%
	 \ifx #1\expandafter \@firstoftwo
	 \else \expandafter \@secondoftwo
	 \fi
	}%
	\providecommand \natexlab [1]{#1}%
	\providecommand \enquote  [1]{``#1''}%
	\providecommand \bibnamefont  [1]{#1}%
	\providecommand \bibfnamefont [1]{#1}%
	\providecommand \citenamefont [1]{#1}%
	\providecommand \href@noop [0]{\@secondoftwo}%
	\providecommand \href [0]{\begingroup \@sanitize@url \@href}%
	\providecommand \@href[1]{\@@startlink{#1}\@@href}%
	\providecommand \@@href[1]{\endgroup#1\@@endlink}%
	\providecommand \@sanitize@url [0]{\catcode `\\12\catcode `\$12\catcode `\&12\catcode `\#12\catcode `\^12\catcode `\_12\catcode `\%12\relax}%
	\providecommand \@@startlink[1]{}%
	\providecommand \@@endlink[0]{}%
	\providecommand \url  [0]{\begingroup\@sanitize@url \@url }%
	\providecommand \@url [1]{\endgroup\@href {#1}{\urlprefix }}%
	\providecommand \urlprefix  [0]{URL }%
	\providecommand \Eprint [0]{\href }%
	\providecommand \doibase [0]{https://doi.org/}%
	\providecommand \selectlanguage [0]{\@gobble}%
	\providecommand \bibinfo  [0]{\@secondoftwo}%
	\providecommand \bibfield  [0]{\@secondoftwo}%
	\providecommand \translation [1]{[#1]}%
	\providecommand \BibitemOpen [0]{}%
	\providecommand \bibitemStop [0]{}%
	\providecommand \bibitemNoStop [0]{.\EOS\space}%
	\providecommand \EOS [0]{\spacefactor3000\relax}%
	\providecommand \BibitemShut  [1]{\csname bibitem#1\endcsname}%
	\let\auto@bib@innerbib\@empty
	\bibitem [{\citenamefont {Vernon}\ and\ \citenamefont {Pedersen}(1968)}]{vernon_relaxation_1968}%
	  \BibitemOpen
	  \bibfield  {author} {\bibinfo {author} {\bibfnamefont {F.~L.}\ \bibnamefont {Vernon}}\ and\ \bibinfo {author} {\bibfnamefont {R.~J.}\ \bibnamefont {Pedersen}},\ }\bibfield  {title} {\bibinfo {title} {Relaxation oscillations in josephson junctions},\ }\href {https://doi.org/10.1063/1.1656649} {\bibfield  {journal} {\bibinfo  {journal} {Journal of Applied Physics}\ }\textbf {\bibinfo {volume} {39}},\ \bibinfo {pages} {2661} (\bibinfo {year} {1968})}\BibitemShut {NoStop}%
	\bibitem [{\citenamefont {Whan}\ \emph {et~al.}(1995)\citenamefont {Whan}, \citenamefont {Lobb},\ and\ \citenamefont {Forrester}}]{whan_effect_1995}%
	  \BibitemOpen
	  \bibfield  {author} {\bibinfo {author} {\bibfnamefont {C.~B.}\ \bibnamefont {Whan}}, \bibinfo {author} {\bibfnamefont {C.~J.}\ \bibnamefont {Lobb}},\ and\ \bibinfo {author} {\bibfnamefont {M.~G.}\ \bibnamefont {Forrester}},\ }\bibfield  {title} {\bibinfo {title} {{Effect of inductance in externally shunted Josephson tunnel junctions}},\ }\href {https://doi.org/10.1063/1.359334} {\bibfield  {journal} {\bibinfo  {journal} {Journal of Applied Physics}\ }\textbf {\bibinfo {volume} {77}},\ \bibinfo {pages} {382} (\bibinfo {year} {1995})}\BibitemShut {NoStop}%
	\bibitem [{\citenamefont {Schmid}(1983)}]{schmid_diffusion_1983}%
	  \BibitemOpen
	  \bibfield  {author} {\bibinfo {author} {\bibfnamefont {A.}~\bibnamefont {Schmid}},\ }\bibfield  {title} {\bibinfo {title} {Diffusion and localization in a dissipative quantum system},\ }\href {https://doi.org/10.1103/PhysRevLett.51.1506} {\bibfield  {journal} {\bibinfo  {journal} {Phys. Rev. Lett.}\ }\textbf {\bibinfo {volume} {51}},\ \bibinfo {pages} {1506} (\bibinfo {year} {1983})}\BibitemShut {NoStop}%
	\bibitem [{\citenamefont {Bulgadaev}(1984)}]{bulgadaev_phase_1984}%
	  \BibitemOpen
	  \bibfield  {author} {\bibinfo {author} {\bibfnamefont {S.~A.}\ \bibnamefont {Bulgadaev}},\ }\bibfield  {title} {\bibinfo {title} {Phase diagram of a dissipative quantum system},\ }\href {http://jetpletters.ru/ps/0/article_19477.shtml} {\bibfield  {journal} {\bibinfo  {journal} {JETP Lett.}\ }\textbf {\bibinfo {volume} {39}},\ \bibinfo {pages} {315} (\bibinfo {year} {1984})}\BibitemShut {NoStop}%
	\bibitem [{\citenamefont {Kuzmin}\ \emph {et~al.}(1991)\citenamefont {Kuzmin}, \citenamefont {Nazarov}, \citenamefont {Haviland}, \citenamefont {Delsing},\ and\ \citenamefont {Claeson}}]{kuzmin_coulomb_1991}%
	  \BibitemOpen
	  \bibfield  {author} {\bibinfo {author} {\bibfnamefont {L.~S.}\ \bibnamefont {Kuzmin}}, \bibinfo {author} {\bibfnamefont {Y.~V.}\ \bibnamefont {Nazarov}}, \bibinfo {author} {\bibfnamefont {D.~B.}\ \bibnamefont {Haviland}}, \bibinfo {author} {\bibfnamefont {P.}~\bibnamefont {Delsing}},\ and\ \bibinfo {author} {\bibfnamefont {T.}~\bibnamefont {Claeson}},\ }\bibfield  {title} {\bibinfo {title} {Coulomb blockade and incoherent tunneling of cooper pairs in ultrasmall junctions affected by strong quantum fluctuations},\ }\href {https://doi.org/10.1103/PhysRevLett.67.1161} {\bibfield  {journal} {\bibinfo  {journal} {Phys. Rev. Lett.}\ }\textbf {\bibinfo {volume} {67}},\ \bibinfo {pages} {1161} (\bibinfo {year} {1991})}\BibitemShut {NoStop}%
	\bibitem [{\citenamefont {Yagi}\ \emph {et~al.}(1997)\citenamefont {Yagi}, \citenamefont {Kobayashi},\ and\ \citenamefont {Ootuka}}]{yagi_phase_1997}%
	  \BibitemOpen
	  \bibfield  {author} {\bibinfo {author} {\bibfnamefont {R.}~\bibnamefont {Yagi}}, \bibinfo {author} {\bibfnamefont {S.-i.}\ \bibnamefont {Kobayashi}},\ and\ \bibinfo {author} {\bibfnamefont {Y.}~\bibnamefont {Ootuka}},\ }\bibfield  {title} {\bibinfo {title} {Phase diagram for superconductor-insulator transition in single small josephson junctions with shunt resistor},\ }\href {https://doi.org/10.1143/JPSJ.66.3722} {\bibfield  {journal} {\bibinfo  {journal} {Journal of the Physical Society of Japan}\ }\textbf {\bibinfo {volume} {66}},\ \bibinfo {pages} {3722} (\bibinfo {year} {1997})}\BibitemShut {NoStop}%
	\bibitem [{\citenamefont {Penttil\"a}\ \emph {et~al.}(1999)\citenamefont {Penttil\"a}, \citenamefont {Parts}, \citenamefont {Hakonen}, \citenamefont {Paalanen},\ and\ \citenamefont {Sonin}}]{penttila_superconductor-insulator_1999}%
	  \BibitemOpen
	  \bibfield  {author} {\bibinfo {author} {\bibfnamefont {J.~S.}\ \bibnamefont {Penttil\"a}}, \bibinfo {author} {\bibfnamefont {U.}~\bibnamefont {Parts}}, \bibinfo {author} {\bibfnamefont {P.~J.}\ \bibnamefont {Hakonen}}, \bibinfo {author} {\bibfnamefont {M.~A.}\ \bibnamefont {Paalanen}},\ and\ \bibinfo {author} {\bibfnamefont {E.~B.}\ \bibnamefont {Sonin}},\ }\bibfield  {title} {\bibinfo {title} {``superconductor-insulator transition'' in a single josephson junction},\ }\href {https://doi.org/10.1103/PhysRevLett.82.1004} {\bibfield  {journal} {\bibinfo  {journal} {Phys. Rev. Lett.}\ }\textbf {\bibinfo {volume} {82}},\ \bibinfo {pages} {1004} (\bibinfo {year} {1999})}\BibitemShut {NoStop}%
	\bibitem [{\citenamefont {Murani}\ \emph {et~al.}(2020)\citenamefont {Murani}, \citenamefont {Bourlet}, \citenamefont {le~Sueur}, \citenamefont {Portier}, \citenamefont {Altimiras}, \citenamefont {Esteve}, \citenamefont {Grabert}, \citenamefont {Stockburger}, \citenamefont {Ankerhold},\ and\ \citenamefont {Joyez}}]{murani_absence_2020}%
	  \BibitemOpen
	  \bibfield  {author} {\bibinfo {author} {\bibfnamefont {A.}~\bibnamefont {Murani}}, \bibinfo {author} {\bibfnamefont {N.}~\bibnamefont {Bourlet}}, \bibinfo {author} {\bibfnamefont {H.}~\bibnamefont {le~Sueur}}, \bibinfo {author} {\bibfnamefont {F.}~\bibnamefont {Portier}}, \bibinfo {author} {\bibfnamefont {C.}~\bibnamefont {Altimiras}}, \bibinfo {author} {\bibfnamefont {D.}~\bibnamefont {Esteve}}, \bibinfo {author} {\bibfnamefont {H.}~\bibnamefont {Grabert}}, \bibinfo {author} {\bibfnamefont {J.}~\bibnamefont {Stockburger}}, \bibinfo {author} {\bibfnamefont {J.}~\bibnamefont {Ankerhold}},\ and\ \bibinfo {author} {\bibfnamefont {P.}~\bibnamefont {Joyez}},\ }\bibfield  {title} {\bibinfo {title} {Absence of a dissipative quantum phase transition in josephson junctions},\ }\href {https://doi.org/10.1103/PhysRevX.10.021003} {\bibfield  {journal} {\bibinfo  {journal} {Phys. Rev. X}\ }\textbf {\bibinfo {volume} {10}},\ \bibinfo {pages} {021003} (\bibinfo {year} {2020})}\BibitemShut {NoStop}%
	\bibitem [{\citenamefont {Kuzmin}\ \emph {et~al.}(2023)\citenamefont {Kuzmin}, \citenamefont {Mehta}, \citenamefont {Grabon}, \citenamefont {Mencia}, \citenamefont {Burshtein}, \citenamefont {Goldstein},\ and\ \citenamefont {Manucharyan}}]{kuzmin_observation_2023}%
	  \BibitemOpen
	  \bibfield  {author} {\bibinfo {author} {\bibfnamefont {R.}~\bibnamefont {Kuzmin}}, \bibinfo {author} {\bibfnamefont {N.}~\bibnamefont {Mehta}}, \bibinfo {author} {\bibfnamefont {N.}~\bibnamefont {Grabon}}, \bibinfo {author} {\bibfnamefont {R.~A.}\ \bibnamefont {Mencia}}, \bibinfo {author} {\bibfnamefont {A.}~\bibnamefont {Burshtein}}, \bibinfo {author} {\bibfnamefont {M.}~\bibnamefont {Goldstein}},\ and\ \bibinfo {author} {\bibfnamefont {V.~E.}\ \bibnamefont {Manucharyan}},\ }\bibfield  {title} {\bibinfo {title} {Observation of the schmid-bulgadaev dissipative quantum phase transition},\ }\href {https://arxiv.org/abs/2304.05806} {\bibfield  {journal} {\bibinfo  {journal} {arXiv:2304.05806}\ } (\bibinfo {year} {2023})}\BibitemShut {NoStop}%
	\bibitem [{\citenamefont {Morel}\ and\ \citenamefont {Mora}(2021)}]{morel_double-periodic_2021}%
	  \BibitemOpen
	  \bibfield  {author} {\bibinfo {author} {\bibfnamefont {T.}~\bibnamefont {Morel}}\ and\ \bibinfo {author} {\bibfnamefont {C.}~\bibnamefont {Mora}},\ }\bibfield  {title} {\bibinfo {title} {Double-periodic josephson junctions in a quantum dissipative environment},\ }\href {https://doi.org/10.1103/PhysRevB.104.245417} {\bibfield  {journal} {\bibinfo  {journal} {Phys. Rev. B}\ }\textbf {\bibinfo {volume} {104}},\ \bibinfo {pages} {245417} (\bibinfo {year} {2021})}\BibitemShut {NoStop}%
	\bibitem [{\citenamefont {Masuki}\ \emph {et~al.}(2022)\citenamefont {Masuki}, \citenamefont {Sudo}, \citenamefont {Oshikawa},\ and\ \citenamefont {Ashida}}]{masuki_absence_2022}%
	  \BibitemOpen
	  \bibfield  {author} {\bibinfo {author} {\bibfnamefont {K.}~\bibnamefont {Masuki}}, \bibinfo {author} {\bibfnamefont {H.}~\bibnamefont {Sudo}}, \bibinfo {author} {\bibfnamefont {M.}~\bibnamefont {Oshikawa}},\ and\ \bibinfo {author} {\bibfnamefont {Y.}~\bibnamefont {Ashida}},\ }\bibfield  {title} {\bibinfo {title} {Absence versus presence of dissipative quantum phase transition in josephson junctions},\ }\href {https://doi.org/10.1103/PhysRevLett.129.087001} {\bibfield  {journal} {\bibinfo  {journal} {Phys. Rev. Lett.}\ }\textbf {\bibinfo {volume} {129}},\ \bibinfo {pages} {087001} (\bibinfo {year} {2022})}\BibitemShut {NoStop}%
	\bibitem [{\citenamefont {Altimiras}\ \emph {et~al.}(2023)\citenamefont {Altimiras}, \citenamefont {Esteve}, \citenamefont {Çağlar Girit}, \citenamefont {le~Sueur},\ and\ \citenamefont {Joyez}}]{carles_absence_2023}%
	  \BibitemOpen
	  \bibfield  {author} {\bibinfo {author} {\bibfnamefont {C.}~\bibnamefont {Altimiras}}, \bibinfo {author} {\bibfnamefont {D.}~\bibnamefont {Esteve}}, \bibinfo {author} {\bibnamefont {Çağlar Girit}}, \bibinfo {author} {\bibfnamefont {H.}~\bibnamefont {le~Sueur}},\ and\ \bibinfo {author} {\bibfnamefont {P.}~\bibnamefont {Joyez}},\ }\bibfield  {title} {\bibinfo {title} {Absence of a dissipative quantum phase transition in josephson junctions: Theory},\ }\href {https://arxiv.org/abs/2312.14754} {\bibfield  {journal} {\bibinfo  {journal} {arXiv:2312.14754}\ } (\bibinfo {year} {2023})}\BibitemShut {NoStop}%
	\bibitem [{\citenamefont {Averin}\ \emph {et~al.}(1985)\citenamefont {Averin}, \citenamefont {Zorin},\ and\ \citenamefont {Likharev}}]{averin_bloch_1985}%
	  \BibitemOpen
	  \bibfield  {author} {\bibinfo {author} {\bibfnamefont {D.}~\bibnamefont {Averin}}, \bibinfo {author} {\bibfnamefont {A.}~\bibnamefont {Zorin}},\ and\ \bibinfo {author} {\bibfnamefont {K.}~\bibnamefont {Likharev}},\ }\bibfield  {title} {\bibinfo {title} {Bloch oscillations in small josephson junctions},\ }\href {http://www.jetp.ras.ru/cgi-bin/dn/e_061_02_0407.pdf} {\bibfield  {journal} {\bibinfo  {journal} {Sov. Phys. JETP}\ }\textbf {\bibinfo {volume} {61}},\ \bibinfo {pages} {407} (\bibinfo {year} {1985})}\BibitemShut {NoStop}%
	\bibitem [{\citenamefont {Lehtinen}\ \emph {et~al.}(2012)\citenamefont {Lehtinen}, \citenamefont {Zakharov},\ and\ \citenamefont {Arutyunov}}]{lehtinen_coulomb_2012}%
	  \BibitemOpen
	  \bibfield  {author} {\bibinfo {author} {\bibfnamefont {J.~S.}\ \bibnamefont {Lehtinen}}, \bibinfo {author} {\bibfnamefont {K.}~\bibnamefont {Zakharov}},\ and\ \bibinfo {author} {\bibfnamefont {K.~Y.}\ \bibnamefont {Arutyunov}},\ }\bibfield  {title} {\bibinfo {title} {Coulomb blockade and bloch oscillations in superconducting ti nanowires},\ }\href {https://doi.org/10.1103/PhysRevLett.109.187001} {\bibfield  {journal} {\bibinfo  {journal} {Phys. Rev. Lett.}\ }\textbf {\bibinfo {volume} {109}},\ \bibinfo {pages} {187001} (\bibinfo {year} {2012})}\BibitemShut {NoStop}%
	\bibitem [{\citenamefont {Shaikhaidarov}\ \emph {et~al.}(2022)\citenamefont {Shaikhaidarov}, \citenamefont {Kim}, \citenamefont {Dunstan}, \citenamefont {Antonov}, \citenamefont {Linzen}, \citenamefont {Ziegler}, \citenamefont {Golubev}, \citenamefont {Antonov}, \citenamefont {Il'ichev},\ and\ \citenamefont {Astafiev}}]{shaikhaidarov_quantized_2022}%
	  \BibitemOpen
	  \bibfield  {author} {\bibinfo {author} {\bibfnamefont {R.~S.}\ \bibnamefont {Shaikhaidarov}}, \bibinfo {author} {\bibfnamefont {K.~H.}\ \bibnamefont {Kim}}, \bibinfo {author} {\bibfnamefont {J.~W.}\ \bibnamefont {Dunstan}}, \bibinfo {author} {\bibfnamefont {I.~V.}\ \bibnamefont {Antonov}}, \bibinfo {author} {\bibfnamefont {S.}~\bibnamefont {Linzen}}, \bibinfo {author} {\bibfnamefont {M.}~\bibnamefont {Ziegler}}, \bibinfo {author} {\bibfnamefont {D.~S.}\ \bibnamefont {Golubev}}, \bibinfo {author} {\bibfnamefont {V.~N.}\ \bibnamefont {Antonov}}, \bibinfo {author} {\bibfnamefont {E.~V.}\ \bibnamefont {Il'ichev}},\ and\ \bibinfo {author} {\bibfnamefont {O.~V.}\ \bibnamefont {Astafiev}},\ }\bibfield  {title} {\bibinfo {title} {Quantized current steps due to the a.c. coherent quantum phase-slip effect},\ }\href {https://doi.org/10.1038/s41586-022-04947-z} {\bibfield  {journal} {\bibinfo  {journal} {Nature}\ }\textbf {\bibinfo {volume} {608}},\ \bibinfo {pages} {45} (\bibinfo {year} {2022})}\BibitemShut {NoStop}%
	\bibitem [{\citenamefont {Crescini}\ \emph {et~al.}(2023)\citenamefont {Crescini}, \citenamefont {Cailleaux}, \citenamefont {Guichard}, \citenamefont {Naud}, \citenamefont {Buisson}, \citenamefont {W.~Murch},\ and\ \citenamefont {Roch}}]{crescini_evidence_2023}%
	  \BibitemOpen
	  \bibfield  {author} {\bibinfo {author} {\bibfnamefont {N.}~\bibnamefont {Crescini}}, \bibinfo {author} {\bibfnamefont {S.}~\bibnamefont {Cailleaux}}, \bibinfo {author} {\bibfnamefont {W.}~\bibnamefont {Guichard}}, \bibinfo {author} {\bibfnamefont {C.}~\bibnamefont {Naud}}, \bibinfo {author} {\bibfnamefont {O.}~\bibnamefont {Buisson}}, \bibinfo {author} {\bibfnamefont {K.}~\bibnamefont {W.~Murch}},\ and\ \bibinfo {author} {\bibfnamefont {N.}~\bibnamefont {Roch}},\ }\bibfield  {title} {\bibinfo {title} {Evidence of dual shapiro steps in a josephson junction array},\ }\href {https://doi.org/10.1038/s41567-023-01961-4} {\bibfield  {journal} {\bibinfo  {journal} {Nature Physics}\ }\textbf {\bibinfo {volume} {19}},\ \bibinfo {pages} {851} (\bibinfo {year} {2023})}\BibitemShut {NoStop}%
	\bibitem [{\citenamefont {{\AA}gren}\ \emph {et~al.}(2001)\citenamefont {{\AA}gren}, \citenamefont {Andersson},\ and\ \citenamefont {Haviland}}]{agren_kinetic_2001}%
	  \BibitemOpen
	  \bibfield  {author} {\bibinfo {author} {\bibfnamefont {P.}~\bibnamefont {{\AA}gren}}, \bibinfo {author} {\bibfnamefont {K.}~\bibnamefont {Andersson}},\ and\ \bibinfo {author} {\bibfnamefont {D.~B.}\ \bibnamefont {Haviland}},\ }\bibfield  {title} {\bibinfo {title} {Kinetic inductance and coulomb blockade in one dimensional josephson junction arrays},\ }\href {https://doi.org/10.1023/A:1017594322332} {\bibfield  {journal} {\bibinfo  {journal} {Journal of Low Temperature Physics}\ }\textbf {\bibinfo {volume} {124}},\ \bibinfo {pages} {291} (\bibinfo {year} {2001})}\BibitemShut {NoStop}%
	\bibitem [{\citenamefont {Vogt}\ \emph {et~al.}(2015)\citenamefont {Vogt}, \citenamefont {Sch\"afer}, \citenamefont {Rotzinger}, \citenamefont {Cui}, \citenamefont {Fiebig}, \citenamefont {Shnirman},\ and\ \citenamefont {Ustinov}}]{vogt_one-dimensional_2015}%
	  \BibitemOpen
	  \bibfield  {author} {\bibinfo {author} {\bibfnamefont {N.}~\bibnamefont {Vogt}}, \bibinfo {author} {\bibfnamefont {R.}~\bibnamefont {Sch\"afer}}, \bibinfo {author} {\bibfnamefont {H.}~\bibnamefont {Rotzinger}}, \bibinfo {author} {\bibfnamefont {W.}~\bibnamefont {Cui}}, \bibinfo {author} {\bibfnamefont {A.}~\bibnamefont {Fiebig}}, \bibinfo {author} {\bibfnamefont {A.}~\bibnamefont {Shnirman}},\ and\ \bibinfo {author} {\bibfnamefont {A.~V.}\ \bibnamefont {Ustinov}},\ }\bibfield  {title} {\bibinfo {title} {One-dimensional josephson junction arrays: Lifting the coulomb blockade by depinning},\ }\href {https://doi.org/10.1103/PhysRevB.92.045435} {\bibfield  {journal} {\bibinfo  {journal} {Phys. Rev. B}\ }\textbf {\bibinfo {volume} {92}},\ \bibinfo {pages} {045435} (\bibinfo {year} {2015})}\BibitemShut {NoStop}%
	\bibitem [{\citenamefont {Cedergren}\ \emph {et~al.}(2017)\citenamefont {Cedergren}, \citenamefont {Ackroyd}, \citenamefont {Kafanov}, \citenamefont {Vogt}, \citenamefont {Shnirman},\ and\ \citenamefont {Duty}}]{cedergren_insulating_2017}%
	  \BibitemOpen
	  \bibfield  {author} {\bibinfo {author} {\bibfnamefont {K.}~\bibnamefont {Cedergren}}, \bibinfo {author} {\bibfnamefont {R.}~\bibnamefont {Ackroyd}}, \bibinfo {author} {\bibfnamefont {S.}~\bibnamefont {Kafanov}}, \bibinfo {author} {\bibfnamefont {N.}~\bibnamefont {Vogt}}, \bibinfo {author} {\bibfnamefont {A.}~\bibnamefont {Shnirman}},\ and\ \bibinfo {author} {\bibfnamefont {T.}~\bibnamefont {Duty}},\ }\bibfield  {title} {\bibinfo {title} {Insulating josephson junction chains as pinned luttinger liquids},\ }\href {https://link.aps.org/doi/10.1103/PhysRevLett.119.167701} {\bibfield  {journal} {\bibinfo  {journal} {Physical review letters}\ }\textbf {\bibinfo {volume} {119}},\ \bibinfo {pages} {167701} (\bibinfo {year} {2017})}\BibitemShut {NoStop}%
	\bibitem [{\citenamefont {Lotkhov}\ \emph {et~al.}(2007)\citenamefont {Lotkhov}, \citenamefont {Krupenin},\ and\ \citenamefont {Zorin}}]{lotkhov_cooper_2007}%
	  \BibitemOpen
	  \bibfield  {author} {\bibinfo {author} {\bibfnamefont {S.~V.}\ \bibnamefont {Lotkhov}}, \bibinfo {author} {\bibfnamefont {V.~A.}\ \bibnamefont {Krupenin}},\ and\ \bibinfo {author} {\bibfnamefont {A.~B.}\ \bibnamefont {Zorin}},\ }\bibfield  {title} {\bibinfo {title} {Cooper pair transport in a resistor-biased josephson junction array},\ }\href {https://doi.org/10.1109/TIM.2007.890793} {\bibfield  {journal} {\bibinfo  {journal} {IEEE Transactions on Instrumentation and Measurement}\ }\textbf {\bibinfo {volume} {56}},\ \bibinfo {pages} {491} (\bibinfo {year} {2007})}\BibitemShut {NoStop}%
	\bibitem [{\citenamefont {Mukhopadhyay}\ \emph {et~al.}(2023)\citenamefont {Mukhopadhyay}, \citenamefont {Senior}, \citenamefont {Saez-Mollejo}, \citenamefont {Puglia}, \citenamefont {Zemlicka}, \citenamefont {Fink},\ and\ \citenamefont {Higginbotham}}]{mukhopadhyay_superconductivity_2023}%
	  \BibitemOpen
	  \bibfield  {author} {\bibinfo {author} {\bibfnamefont {S.}~\bibnamefont {Mukhopadhyay}}, \bibinfo {author} {\bibfnamefont {J.}~\bibnamefont {Senior}}, \bibinfo {author} {\bibfnamefont {J.}~\bibnamefont {Saez-Mollejo}}, \bibinfo {author} {\bibfnamefont {D.}~\bibnamefont {Puglia}}, \bibinfo {author} {\bibfnamefont {M.}~\bibnamefont {Zemlicka}}, \bibinfo {author} {\bibfnamefont {J.~M.}\ \bibnamefont {Fink}},\ and\ \bibinfo {author} {\bibfnamefont {A.~P.}\ \bibnamefont {Higginbotham}},\ }\bibfield  {title} {\bibinfo {title} {Superconductivity from a melted insulator in josephson junction arrays},\ }\href {https://www.nature.com/articles/s41567-023-02161-w} {\bibfield  {journal} {\bibinfo  {journal} {Nature Physics}\ }\textbf {\bibinfo {volume} {19}},\ \bibinfo {pages} {1630} (\bibinfo {year} {2023})}\BibitemShut {NoStop}%
	\bibitem [{Note1()}]{Note1}%
	  \BibitemOpen
	  \bibinfo {note} {We have checked numerically that this is adequate for small fields}\BibitemShut {NoStop}%
	\bibitem [{\citenamefont {Tinkham}(1996)}]{tinkham_introduction_1996}%
	  \BibitemOpen
	  \bibfield  {author} {\bibinfo {author} {\bibfnamefont {M.}~\bibnamefont {Tinkham}},\ }\href@noop {} {\emph {\bibinfo {title} {Introduction to Superconductivity}}},\ \bibinfo {edition} {2nd}\ ed.\ (\bibinfo  {publisher} {McGraw-Hill, Inc.},\ \bibinfo {year} {1996})\ pp.\ \bibinfo {pages} {118--119, 127--131, 390--393}\BibitemShut {NoStop}%
	\bibitem [{\citenamefont {Wellstood}\ \emph {et~al.}(1989)\citenamefont {Wellstood}, \citenamefont {Urbina},\ and\ \citenamefont {Clarke}}]{wellstood_hot-electron_1989}%
	  \BibitemOpen
	  \bibfield  {author} {\bibinfo {author} {\bibfnamefont {F.~C.}\ \bibnamefont {Wellstood}}, \bibinfo {author} {\bibfnamefont {C.}~\bibnamefont {Urbina}},\ and\ \bibinfo {author} {\bibfnamefont {J.}~\bibnamefont {Clarke}},\ }\bibfield  {title} {\bibinfo {title} {{Hot‐electron limitation to the sensitivity of the dc superconducting quantum interference device}},\ }\href {https://doi.org/10.1063/1.101062} {\bibfield  {journal} {\bibinfo  {journal} {Applied Physics Letters}\ }\textbf {\bibinfo {volume} {54}},\ \bibinfo {pages} {2599} (\bibinfo {year} {1989})}\BibitemShut {NoStop}%
	\bibitem [{\citenamefont {Kauppinen}\ and\ \citenamefont {Pekola}(1996)}]{kauppinen_electron-phonon_1996}%
	  \BibitemOpen
	  \bibfield  {author} {\bibinfo {author} {\bibfnamefont {J.~P.}\ \bibnamefont {Kauppinen}}\ and\ \bibinfo {author} {\bibfnamefont {J.~P.}\ \bibnamefont {Pekola}},\ }\bibfield  {title} {\bibinfo {title} {Electron-phonon heat transport in arrays of al islands with submicrometer-sized tunnel junctions},\ }\href {https://doi.org/10.1103/PhysRevB.54.R8353} {\bibfield  {journal} {\bibinfo  {journal} {Phys. Rev. B}\ }\textbf {\bibinfo {volume} {54}},\ \bibinfo {pages} {R8353} (\bibinfo {year} {1996})}\BibitemShut {NoStop}%
	\end{thebibliography}
%
	
\end{document}